\documentclass[nopubldata]{lpor2014}
\usepackage{blindtext}
\usepackage{hyperref}
\usepackage{ulem}
\hypersetup{%
  colorlinks,
  plainpages=false,
  pdfpagelabels,
  breaklinks,
  pdfview=Fit,
  bookmarksopen,
  bookmarksnumbered,
  linkcolor=black,
  anchorcolor=black,
  citecolor=black,
  filecolor=black,
  urlcolor=LPRred
  }%
\graphicspath{{./figs/}}
\setpages[]{}
\setvolume[0]{0}
\setyear{2011}%
\setdoi{201100000}%
\oddheadlogotext{}{Guide}
\begin{document}

\titlefigure{fig0}
\abstract{%
   Nonlinear wave-mixing in mesoscopic silicon structures is a fundamental nonlinear process with broad impact and applications. Silicon nanowire waveguides, in particular, have large third-order Kerr nonlinearity, enabling salient and abundant four-wave mixing dynamics and functionalities. Besides Kerr effect, in silicon waveguide two-photon absorption generates high free-carrier densities, with corresponding fifth-order nonlinearity in the forms of free-carrier dispersion and free-carrier absorption. However, whether these fifth-order free carrier nonlinear effects can lead to six-wave mixing dynamics still remains an open question until now. Here, we report the first demonstration of free-carrier induced six-wave mixing in silicon nanowires. Unique features including inverse detuning dependence of six-wave mixing efficiency, and its higher sensitivity to pump power, are originally observed and verified by analytical prediction and numerical modeling. Additionally, asymmetric sideband generation is observed for different laser detunings, resulting from the phase-sensitive interactions between free-carrier six-wave mixing and Kerr four-wave mixing dynamics. These discoveries provide a new path for nonlinear multi-wave interactions in nanoscale platforms.}

\title{Six-wave mixing induced by free-carrier plasma in silicon nanowire waveguides}
\titlerunning{Six-wave mixing induced by free-carrier plasma in silicon nanowire waveguides}

\author{Heng Zhou\inst{1,*}, Mingle Liao\inst{1}, Shu-Wei Huang\inst{2}, Linjie Zhou\inst{3}, Kun Qiu\inst{1} and Chee Wei Wong\inst{2,*}}%
\authorrunning{Heng Zhou, etc.}
\mail{\email{zhouheng@uestc.edu.cn; cheewei.wong@ucla.edu}}

\institute{%
Key Lab of Optical Fiber Sensing and Communication Networks, University of Electronic Science and Technology of China, Chengdu 611731, China.\\
\and
Mesoscopic Optics and Quantum Electronics Laboratory, University of California, Los Angeles, CA 90095, USA.\\
\and
State Key Laboratory of Advanced Optical Communication Systems and Networks, Shanghai Jiao Tong University, Shanghai 200240, China.
}


\maketitle

\section{Introduction}
\label{sec:intro}

Silicon solid-state nanophotonic structures have a large third-order Kerr nonlinearity and strong light confinement, enabling nonlinear optic dynamics with broad impact and applications. Four-wave mixing (FWM), as an elemental nonlinear process, has been deeply investigated in nanoscale silicon platforms [1-5] and implemented in a multitude of functionalities, ranging from optical signal regeneration [6], mid-infrared frequency conversion [7-8], phase conjugation [9], continuum generation [10], regenerative oscillations [11], correlated photons generation [12], and spectroscopy [13]. Essentially, FWM arises when two intense laser fields cause oscillation of the refractive index via the Kerr effect, which in turn imposes nonlinear phase modulation back onto the input driving fields themselves, producing modulation sidebands at new frequencies that satisfy photon energy and momentum conservation conditions [14, 15]. In silicon nonlinear waveguide, in addition to the Kerr effect, two-photon absorption (TPA) generates considerable free-carrier densities, with corresponding nonlinearity and change of refractive index via free-carrier dispersion (FCD) and free-carrier absorption (FCA) [14]. Importantly, the generation of free-carrier density via TPA is already quadratically proportional to the incident laser intensity, making the cascaded refractive index modulation via FCA/FCD a fifth-order nonlinear dynamics on top of the third-order Kerr nonlinearity [16-24]. Various FCD/FCA induced nonlinear phenomena have been demonstrated in silicon, such as soliton fission [18, 19], soliton compression [20], frequency shift [21], and spectrum broadening [23, 24]. However, a fundamentally important question---akin to third-order Kerr effect generating FWM, whether the fifth-order FCD/FCA can give rise to six-wave mixing dynamics---has not been probed until now.

Here we present the first demonstration of free-carrier induced six-wave mixing (FC-SWM) in silicon waveguide. We show a non-dispersion-induced inverse dependence of the FC-SWM strength to input laser detunings, which confirms the existence of FC-SWM resulting in the predominance of FC-SWM over FWM at small laser detunings. Furthermore, we map out the stronger dependence of FC-SWM to input pump power compared to conventional Kerr FWM. Finally, we observe asymmetric sideband generation efficiencies and identify the phase sensitive interaction between FC-SWM and FWM as the mechanism for such symmetry breaking.

\section{Concept and theoretical analysis}
\label{sec:markupcmd}

Kerr and free-carrier nonlinear dynamics in silicon can be described by the nonlinear Schr\"{o}dinger equation (NLSE), which when coupled with the free-carrier generation and recombination dynamics are governed by [14]:

\begin{equation}
  \label{eq:useless}
\begin{array}{l}
\frac{{\partial E}}{{\partial z}}{\rm{ = }} - i\frac{{{\beta _2}}}{2}\frac{{{\partial ^2}E}}{{\partial {t^2}}} + \frac{{{\beta _3}}}{6}\frac{{{\partial ^3}E}}{{\partial {t^3}}} - \frac{{{\alpha _l}}}{2}E\\\\
\;\;\;\;\;\;\;\;\; + (i{\gamma} - \frac{{{\beta _{TPA}}}}{{2{A_0}}}){\left| E \right|^2}E - (i\delta  + \sigma ){N_c}E
\end{array}
\end{equation}

\begin{equation}
  \label{eq:useless}
  \begin{array}{l}
{N_c} = \int_0^t {\left( {\frac{{{\beta _{TPA}}}}{{2A_0^2h{v_0}}}{{\left| {E(z,\tau )} \right|}^4} - \frac{{{N_c}}}{{{\tau _c}}}} \right)d\tau }
\end{array}
\end{equation}

In Eq. (1-2), $E$ is the slowly-varying envelope of the overall input fields into the nanowire waveguide, $\beta_n$ is the $n$th order dispersion, $\gamma$ is the effective Kerr nonlinear coefficient, $\beta_{TPA}$ is the degenerate TPA coefficient, $N_c$ is the free carrier density, $\delta$ and $\sigma$ are the Drude FCD and FCA coefficients respectively, $A_0$ denotes the effective mode area, $\tau_c$ is the free-carrier lifetime, $h$ is the Plank's constant, and $v_0$ is the pump frequency. To analytically derive the wave mixing process, the input light field $E$ is described by ${{A_1}\cos ({\omega _1}t) + {A_2}\cos ({\omega _2}t)}$. As shown in Fig. 1(a), here we utilize two input laser frequencies to study the degenerate FWM via ${\chi ^{(3)}}\left( {2{\omega _1} - {\omega _2};{\omega _1},{\omega _1},{\omega _2}} \right)$ and the degenerate FC-SWM via ${\chi ^{(5)}_{FC}}\left( {2{\omega _1} - {\omega _2};{\omega _1},{\omega _1},{-\omega _1}, {\omega _1}, {\omega _2}} \right)$. Such two-frequency configuration can greatly reduce the complexity of theoretical derivation without loss of generality (more discussions detailed in Supporting Information [25]). Meanwhile, it should be noted that, the ${\chi ^{(5)}_{FC}}$ is induced by FCD/FCA, and has no contribution from the fifth-order electronic nonlinearity $n_4$ [14]. Considering the evolution of the input field and neglecting the dispersion of the waveguide, the nonlinear wave mixing strength ($NM$) experienced by the input field $E$ can be written as (detailed derivation in Supporting Information [25]):

\begin{equation}
  \label{eq:useless}
\begin{array}{l}
NM = \exp \left( {iG\cos (bt) + \frac{{iD}}{b}\sin (bt)} \right)\\\\
\;\;\;\;\;\;\;\; \;\;\; \;   \times \left( {1 + P\cos (bt)} \right) \times \left( {1 + \frac{A}{b}\sin (bt)} \right)
\end{array}
\end{equation}

Here $b={\omega_1} - {\omega_2}$ is the frequency detuning between input lasers, $P=-L{A_1}{A_2}{\beta _{TPA}}{(2{A_0})^{-1}}$, $G=L{A_1}{A_2}{\gamma}$, $A = D\sigma {\delta ^{-1}}$, $D=-3LA_1^3{A_2}\delta {\beta_{TPA}}{(4{A_0}^2h{v_0}) ^{-1}}$, and $L$ is the waveguide length.Herein $G$, $P$, $D$, $A$ represent the effects of Kerr, TPA, FCD, and FCA, respectively. In our study we consider that TPA, FCA, and FCD responds instantaneously to the beating oscillation corresponding to the input laser detunings tested in the experiments ($b/2\pi<$1 THz) [26]. Importantly, Eq. (3) shows that the Kerr and FCD effects cause nonlinear phase modulations that have a $\pi/2$ phase offset with respect to each other (the first and second terms on the right-hand-side of Eq. (3) respectively), while TPA and FCA cause nonlinear intensity modulations that also have a $\pi/2$ phase offset (the third and fourth terms on the right-hand-side of Eq. (3) respectively). Furthermore, Eq. (3) can be re-written as:

\begin{equation}
  \label{eq:useless}
NM = {\rm{exp}}({iH\sin(bt+\theta)})\times({1+M{\rm{cos}}(bt+\psi )})
\end{equation}

Therein, $H=\sqrt{{(D/b}^2)+ {G^2}}$, $M=\sqrt{(A/b)^2+P^2}$, $\theta=\arctan({Gb/D} )$, $\pi/2\leqslant\theta<\pi$, and $\psi=\arctan(-A/Pb)$, $\pi/2\leqslant\psi<\pi$. The values of $\theta$ and $\psi$ are determined by the signs of $G$, $P$, $D$ and $A$. Using the Bessel expansion, we thus arrive at:

\begin{equation}
  \label{eq:useless}
\begin{array}{l}
NM = \sum\limits_{n =  - 1}^1 {{J_n}\left( H \right){e^{inbt + in\theta }}} \\\\
\;\;\;\;\;\;\;\;\;\; \;\; \; \times \left( {1 + \frac{1}{2}M\left( {{e^{ibt + i\psi }} + {e^{ - ibt - i\psi }}} \right)} \right)
\end{array}
\end{equation}

In Eq. (5) $J_n$ is the $n$th order Bessel function. Simplifying Eq. (5) with frequency detunings $\pm b$ respectively and we finally obtain [25]:

\begin{equation}
  \label{eq:useless}
N{M_b} = \;\frac{{\rm{1}}}{{\rm{2}}}M{J_0}{e^{i\psi }} + {J_1}{e^{i\theta }}
\end{equation}

\begin{equation}
  \label{eq:useless}
N{M_{ - b}} = \;\frac{{\rm{1}}}{{\rm{2}}}M{J_0}{e^{ - i\psi }} - {J_1}{e^{ - i\theta }}
\end{equation}

First, as seen in Eq. (5) and the expressions of $H$ and $M$, the FC-SWM components induced by FCD and FCA have a inverse dependence on the input laser detuning ($1/b$), while FWM from Kerr and TPA is detuning independent. Second, the effective fifth-order FC-SWM components are proportional to $A_1^3{A_2}$ (seen in the expressions of $D$ and $A$), while the third-order FWM components are only proportional to $A_1{A_2}$ (seen in the expressions of $G$ and $P$), implying that FC-SWM has a higher sensitivity than FWM. Third, comparing Eq. (6) and (7), under the combinational effects of Kerr, TPA, FCA and FCD in silicon, the generated overall wave mixing sideband power (${\left| {N{M_{ \pm b}}} \right|^2}$) can be asymmetric for positive and negative input laser detunings, even when the waveguide dispersion is neglected. The exact extent of the asymmetry depends on the parameters of the measured waveguide and input light fields. We next present detailed FC-SWM and FWM measurements to demonstrate and validate these theoretical predications.

\begin{figure*}[t]
  \sidecaption
  \includegraphics*[width=0.88\textwidth]{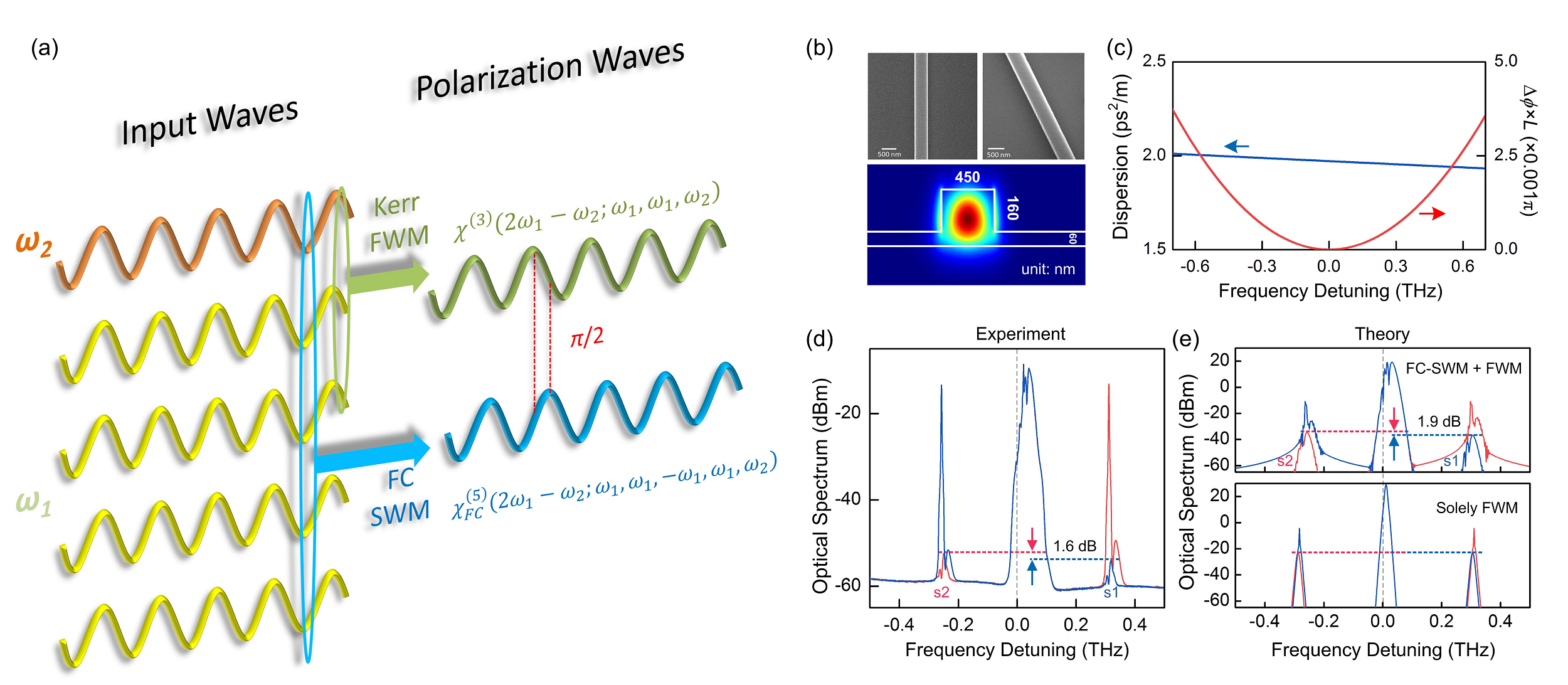}
  \caption{Six-wave mixing spectra and nanowire characteristics. (a) The origin of Kerr FWM and FC-SWM in silicon nanowire waveguide with two input laser frequencies. (b) Scanning electron micrographs of the measured nanowire (upper panel) and simulated mode profile of the fundamental TE$_{11}$ mode using finite-difference time-domain based mode-solver (lower panel). (c) Nanowire dispersion of the TE$_{11}$ mode (left $y$-axis) and the corresponding linear phase mismatch (right $y$-axis). (d) Measured wave mixing spectra with ± 0.28 THz detuning, and the generated sideband idler powers at smaller (s1) and bigger (s2) frequencies. (e) Numerically simulated wave mixing spectra using the NLSE model with experimental parameters, for the case of solely FWM (lower) and Drude FC-SWM plus FWM (upper).The spectrum power discrepancies between numerical and experimental results are due to that the experimental pulse powers were averaged by the 50 MHz pulse repetition rate.}
  \label{fig-2}
\end{figure*}

\section{Experimental results and discussion}
\label{ssec:preamble}

We study a 0.3-mm long rib silicon nanowire fabricated on a silicon-on-insulator wafer, with the scanning electron micrograph shown in Fig. 1(b). Its numerically estimated mode profile and corresponding dispersion of the fundamental TE$_{11}$ mode are shown in Fig. 1(c). The measured nanowire waveguide has a linear loss $\alpha_l=2$ dB/cm, with an effective mode area $A_0=1.3\times10^{-13}$ m$^2$, a degenerate TPA coefficient $\beta_{TPA}=9\times10^{-12}$ m/W, a FCA coefficient $\theta=1.45\times10^{-21} $m$^2$, a FCD coefficient $\delta=1.0\times10^{20} $m$^2$, and a free-carrier recombination lifetime $\tau_c=500$ ps [14, 21-22]. The two incident drive lasers consist of a 50 MHz repetition rate, 100 ps pulsewidth pump field (amplitude $A_1$ and central angular frequency $\omega_1$) and a continuous-wave (c.w.) signal field (amplitude $A_2$ and angular frequency $\omega_2$). The pump pulse train of maximum intra-waveguide peak powers at 11.7 W generates high free-carrier densities on the order of $10^{19}$ cm$^{-3}$. The 50 MHz pulse repetition rate allows sufficient time (more than $40\tau_c$) for complete free-carrier recombination relaxation so that no inter-pulse interference occurs. Moreover, with the small waveguide dispersion and short waveguide length, the dispersion induced linear phase mismatch $\triangle\phi$ is negligibly small within the examined 1550 nm to 1562 nm wavelength range: the maximum value of $\triangle\phi\times L$ is only $0.0037\pi$, as shown in Fig. 1(c), which does not have appreciable impact on the wave mixing process [25, 27-29].

Fig. 1(d) shows two examples of the wave mixing spectra generated in the silicon nanowire waveguide, with input laser detunings equal to $\pm0.28$ THz, and the generated sidebands are denoted as s1 and s2, respectively. Here the input pulse peak power is 11.7 W, and the c.w. power is 0.4 mW. Significantly we observe that the output pulse spectra have apparent FCD-induced spectral blue-shift ($\sim$0.03 THz) [20-24], and the output c.w. spectra exhibit the feature of FCD-induced cross-phase modulation from the pulse [28]. Consequently, the generated wave mixing idler spectra show complex and broadened structures. All these evidences indicate that the measured wave mixing processes are conducted in a regime with strong nonlinear free-carrier dynamics. To confirm this, Fig. 1(e) upper panel shows the numerically simulated wave mixing spectrum using the NLSE model given by Eq. (1-2), which illustrates remarkable agreement with the measurements without fitting. Meanwhile, we note that when the free-carrier dynamics are eliminated from the model, as shown in Fig. 1(e) lower panel, the modeled output spectra lose all the salient features (lineshape broadening, asymmetry, and spectral dips) observed in our measurements.

\subsection{Inverse dependence of FC-SWM on input pump-signal laser detuning.}
\label{ssec:preamble}
From the expressions of parameters $H$ and $M$ in Eq. (4-5), it is noted that the FC-SWM sideband power inversely depends on the input laser detuing ($1/b$). Intrinsically, the $1/b$ factor originates from the temporal integration of $|E|^4$ in the rate equation of carrier density $N_c$, that is, the slower beating oscillation between the input lasers gives rise to higher carrier density fluctuations, and potentially imposes larger six-wave mixing strengths [14]. To demonstrate such dynamics, we scan the input laser detuning from $-$0.7 THz to 0.7 THz by changing the input c.w. frequency and record the wave mixing sideband powers at each scan point, as shown in Fig. 2(a). Intriguingly we observe that for 11.7 W input pulse, as the input laser detuning changes from $\pm0.3$ THz to zero, the generated wave mixing sideband power exhibits an increase of about 3.0 dB, clearly verifying the $1/b$-dependence prediction. To support the measurements, Fig. 2(b) shows the corresponding numerical simulation via the NLSE model Eq. (1-2), as well as the analytical calculation via Eq. (5), both achieving remarkable agreements with our measurements. For comparison, in Fig. 2(b) we plot the theoretical FWM sideband powers induced solely by Kerr effect and TPA, which shows no detuning dependence, and confirms that the waveguide dispersion is negligible compared to the intrinsic $1/b$-dependence of FC-SWM [25]. Particularly, it is seen from Fig. 2(b) that, when laser detuning approaches zero, FC-SWM becomes dominant over FWM (e.g., at 0.1 THz detuning, the FC-SWM sideband is $-$24.0 dBm, about 4.0 dB larger than FWM), strongly supporting the existence of FC-SWM in silicon.

\begin{figure}[t]
  \includegraphics*[width=\linewidth]{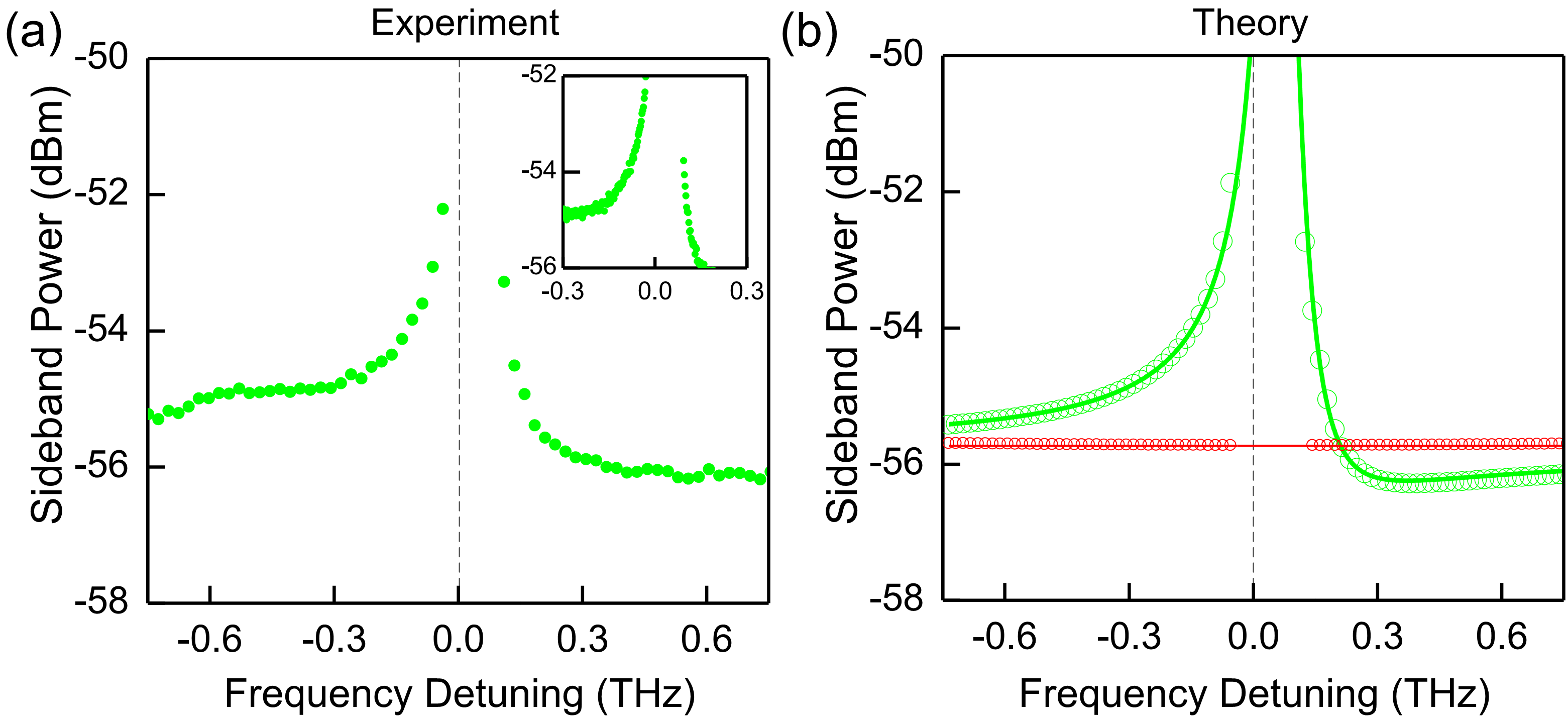}
  \caption{Direct detuning-dependence of the Drude FC-SWM. (a) Experimentally measured wave mixing sideband powers as a function of input laser detuning. The gap in the middle detuning arises because, in such region, the generated sideband components are covered by the pulse spectrum. The laser scanning step is 25 GHz. The inset shows a more densely recorded data with 2.5 GHz step within the closer detuning window from $-$0.3 THz to 0.3 THz. (b) Numerical (open-circle) and analytical (solid-line) wave mixing sideband powers calculated as a function of laser detuning. The green circles and line are for FC-SWM and the red circles and line are for pure FWM (when without the free-carrier contributions). To compare theory with experiment, the calculated sideband powers are down-shifted by 23 dB to compensate for the 50 MHz pulse repetition rate and power attenuation in the measurement.}
  \label{fig-1}
\end{figure}

\subsection{Strong pump power dependency of FC-SWM.}
\label{ssec:preamble}

From Eq. (5) and the expressions of $P$, $G$, $D$, $A$, the FWM strength caused by third-order Kerr and TPA is proportional to $A_1A_2$, while the FC-SWM strength induced by the FCD/FCA is proportional to $A_1^3A_2$, rooting in the fifth-order nonlinear property of FC-SWM. To observe this higher pump power sensitivity of FC-SWM than FWM, Fig. 3(a-b) plots the experimental and theoretical power transfer functions between the input pulse peak power and the generated wave mixing sideband power, under three different frequency detuning values: 0.025 THz, 0.25 THz, and 1.25 THz. Particularly, with the $1/b$ scaling of the FC-SWM demonstrated above, as the detuning decreases from 1.25 THz to 0.025 THz, the contribution of FC-SWM in the overall wave mixing process is substantially enhanced. Simultaneously the slope (linearly fitted) of the power transfer function increases from 0.21 to 1.05, clearly illustrating the stronger pump power dependence of FC-SWM over FWM. Ideally the FC-SWM (FWM) produces a power transfer function with slope equal to 4 (2) [14, 15], but with the concurrent existence of TPA and FCA, the pulse and c.w. laser fields are heavily attenuated in the nanowire waveguide [21-22], and resultantly the measured and calculated power transfer functions are much less steep, as shown in Fig. 3(a-b). Even so, the power transfer function dominated by FC-SWM shows a $5\times$ slope improvement over solely FWM, further confirming the fifth-order nature of FC-SWM.

\begin{figure}[b]
  \includegraphics*[width=\linewidth]{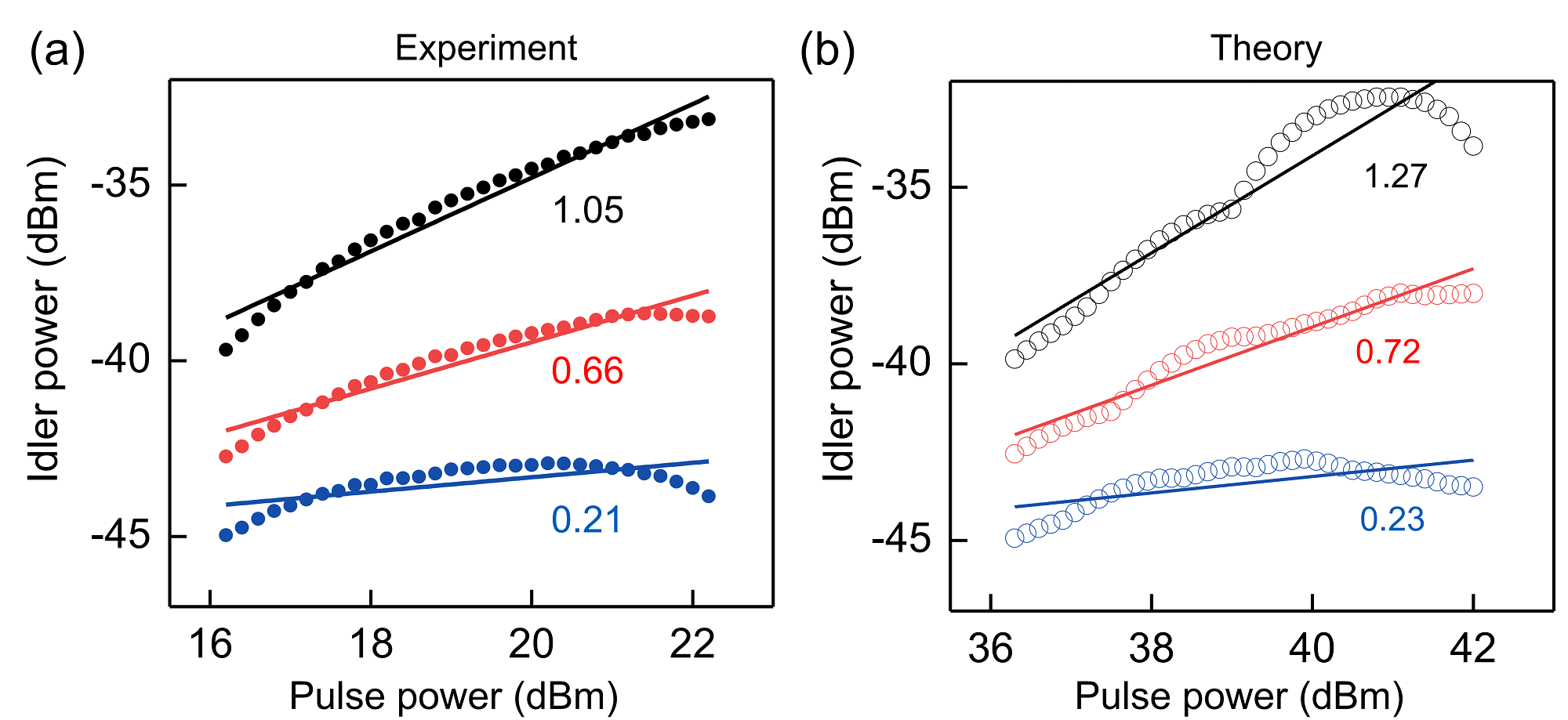}
  \caption{Strong dependences of FC-SWM to pump power. (a) Experimentally measured power transfer function between the input pulse and generated wave mixing sideband, under three different detunings: 0.025 THz (black); 0.25 THz (red); and 1.25 THz (blue). The solid lines are linear fits of the experiment data, with the fitted slope coefficients listed below each corresponding line. (b) Numerically simulated power transfer functions corresponding to the results in panel (a).}
  \label{fig-1}
\end{figure}

\subsection{Phase-sensitive interaction between FC-SWM and FWM.}
\label{ssec:preamble}

It is observed from Fig. 2(a-b) that, with the utilized experimental parameters, FC-SWM and FWM coexist and have comparative magnitudes, which allows us to explore the interplay between six- and four-wave mixing in silicon. As illustrated in Eq. (6) and (7), the combinational effects of FC-SWM and FWM produces an overall wave mixing strength that is asymmetrical for positive and negative input laser detunings. Since the waveguide dispersion is neglected in the derivation, such unconventional asymmetry could arises from the interplay between FC-SWM and FWM [30-31]. The same dynamics is observed experimentally in Fig. 1(d), for two opposite detuning values $\pm$0.28 THz, the generated power of sideband s1 and s2 has an appreciable difference of 1.6 dB. More generally, as observed in Fig. 2(a), the recorded sideband powers exhibit as an apparently asymmetric lineshape as the wavelength detunes from $-$0.70 THz to 0.70 THz.

\begin{table}[t]
  \centering
  \caption{Wave mixing coefficients induced by different constituents of nonlinear processes.}
  \label{tlab}
  \begin{tabular}{@{}lcccc@{}}
    \toprule
    \textbf{Effects} & $NM_{-b}$ & $NM_{b}$\\
    \midrule
    Kerr+TPA & $-MJ_0/2+iJ_1$ & $-MJ_0/2+iJ_1$\\
    TPA+FCA & $Me^{i\psi /2}$ & $Me^{-i\psi /2}$\\
    Kerr+FCD & $-J_1e^{-i\theta}$ & $J_1e^{i\theta}$\\
    FCA+FCD & $-iMJ_0/2+J_1$ & $iMJ_0/2-J_1$\\
    Kerr+FCA & $-iMJ_0/2+iJ_1$ & $iMJ_0/2+iJ_1$\\
    TPA+FCD & $-MJ_0/2+J_1$ & $-MJ_0/2-J_1$\\
    \bottomrule
  \end{tabular}
\end{table}

\begin{figure}[t]
  \includegraphics*[width=\linewidth]{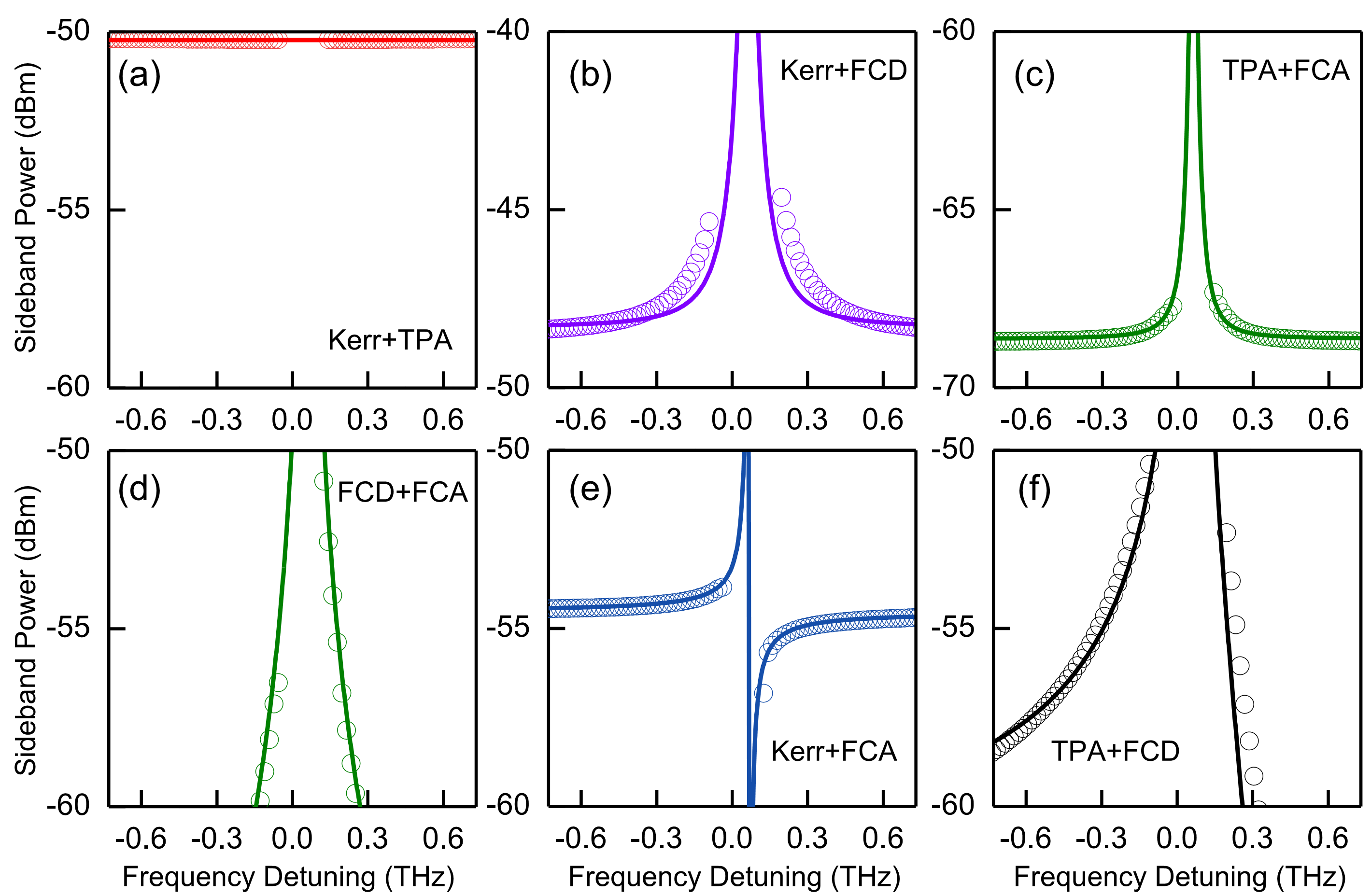}
  \caption{Comparative detuning lineshape symmetries and asymmetries for different nonlinear constituents, noted in Table 1. The open circles are from numerical NLSE simulation and the solid lines are from analytical calculations, based on parameters used from Fig. 2 at 11.7 W. (a) Solely Kerr and TPA without free-carrier dynamics, with negligible detuning dependences. (b) Kerr and FCD constituents, with symmetric lineshape. (c) TPA and FCA constituents, with symmetric lineshape. (d) FCD and FCA constituents, with symmetric lineshape. (e) Kerr and FCA constituents, with lineshape asymmetry. (f) TPA and FCD constituents, with lineshape asymmetry. Note that the numerical results of the sideband powers close to zero detuning are subtracted due to the overlap with the pump spectra linewidth itself.}
  \label{fig-1}
\end{figure}

To elucidate such unconventional sideband evolutions and probe the interaction between different nonlinear wave mixing processes, we tailored Eq. (6-7) with different sets of nonlinear process combinations. As summarized in Table 1, for TPA and FCD, Kerr and FCA, we indeed obtain unequal wave mixing intensity with $\pm b$ detuning; while the other combinations all generate symmetric sideband powers. Importantly, comparing Table 1 and Eq. (3), we conclude that the breaking of wave mixing symmetry can only originate from the interplay between the nonlinear amplitude modulations and the nonlinear phase modulations that have a phase offset of $\pi/2$ (i.e., between TPA and FCD, Kerr and FCA). Fig. 4 shows the analytically calculated wave mixing sideband powers as the function of detuning, and the features predicted in Table 1 are clearly illustrated. To further confirm our analysis, Fig. 4 also presents the numerically simulated sideband powers under different nonlinear effects in Eq. (1-2), which agrees very well with the analytical results.

\begin{figure}[t]
  \includegraphics*[width=\linewidth]{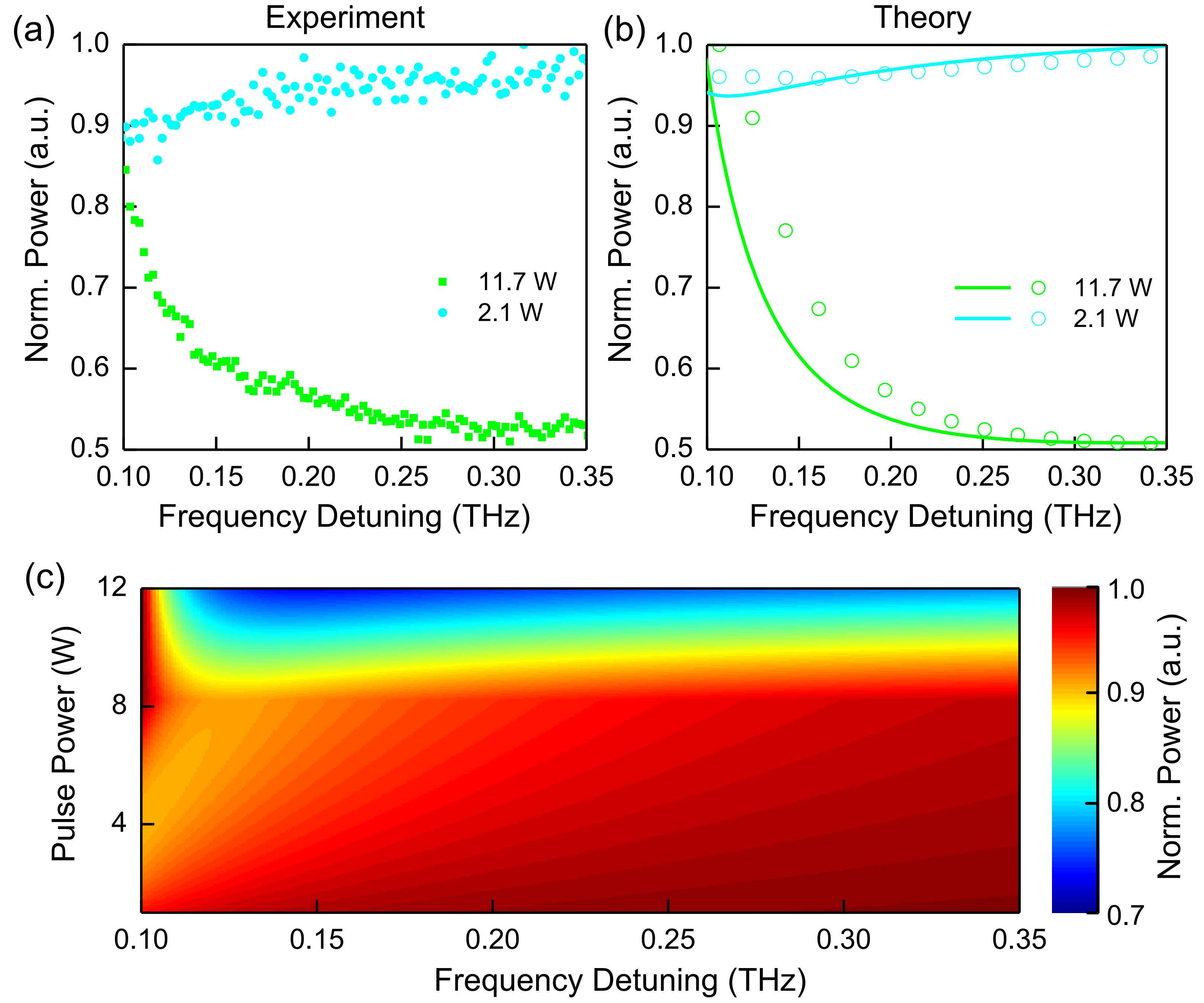}
  \caption{Wave mixing evolution via the phase-sensitive superposition of FC-SWM and FWM. (a) Experimentally measured wave mixing sideband powers as a function of input laser detuning, under two different input pulse peak powers: 11.7 W (green) and 2.1 W (cyan). The laser scanning resolution is 2.5 GHz. (b) Numerical (open-circles) and analytical (solid-lines) calculated wave mixing sideband power evolution, corresponding to a 11.7 W and a 2.1 W pulse. (c) Calculated false-color image of overall wave mixing sideband powers while sweeping input pulse powers from 1 W to 12 W. The sideband powers are normalized to the maximum value for each input pulse power.}
  \label{fig-1}
\end{figure}

Moreover, we find that the phase-sensitive interaction between FC-SWM and FWM significantly modifies the overall sideband generation and opens up new possibilities to manipulate the multi-wave energy exchange in silicon. As indicated in Eq. (6-7), the contributions of FC-SWM and FWM are intertwined within the functions $H$, $M$, $J_0$, $J_1$, which have different monotonicities. Hence the change of each effect can result in variation of the overall sideband power evolution in a non-explicit fashion. To demonstrate this, Fig. 5(a) and 5(b) show the measured and calculated sideband power evolution for two input pulse peak powers. Particularly, for the 11.7 W pulses, FC-SWM dominates FWM such that the sideband evolution approximately follows the $1/b$-dependence featured by FC-SWM, as discussed above in Fig. 2(a-b). On the other hand, when the input pulse power is decreased to 2.1 W, FC-SWM is subjected to more degradation due to its higher dependence on the pump power. Consequently, at this power level, FWM now competes with FC-SWM and results in approximately independence of sideband power to detuning $b$, as shown in Fig. 5(a) and 5(b). More generally, Fig. 5(c) shows the calculated sideband power evolution while sweeping the input pulse power. We observe that as the pulse power increases from 1 W to 12 W, the sideband evolution changes significantly, providing abundant and readily accessible power transfer functions. Such coupled and controllable sideband evolutions can be applicable for all-optical signal processing applications such as all-optical signal regeneration and frequency conversion.

\section{Conclusion}
\label{ssec:preamble}
Here we report the original demonstrations and analysis of Drude free-carrier plasma induced six-wave mixing in silicon nanowire waveguides. Unique features of FC-SWM have been experimentally observed and discussed in-depth. First, the non-dispersion-induced inverse dependence of FC-SWM frequency conversion to the input laser detuning is observed, with FC-SWM sideband power rapidly increasing by 3.0 dB within a 0.3 THz detuning window. Second, the strong dependence of the FC-SWM to input pump powers is illustrated. Third, the phase-sensitive interaction between FC-SWM and FWM is demonstrated for the first time, giving rise to the asymmetric lineshape of sideband power evolution as a function of laser detuning. These observations not only advance our understanding of free-carrier nonlinear dynamics in the multiple-wave regime, but also open up new possibilities for applications based on wave mixing, such as on-chip spectral broadening and all-optical signal processing. Finally, the processes and phenomena demonstrated here can potentially be observed in other physical systems involving plasma nonlinearity, such as gas photoionization in hollow-core photonic crystal fibers [32-34] and light-plasma interactions in semiconductor photonic crystals [35].

\begin{acknowledgement}
  This work was funded by the China 863 Grant 2013AA014402, UESTC young faculty award ZYGX2015-KYQD051, NSF CBET-1438147 and DMR-1611598, Air Force Office of Scientific Research Young Investigator Award (S. W. Huang) FA9550-15-1-0081, and Office of Naval Research N00014-14-1-0041. The authors acknowledge discussions with Xingyu Zhou, Feng Wen, Baojian Wu, and Jinghui Yang.
\end{acknowledgement}

\end{document}